\documentstyle[12pt,aaspp4,psfig]{article}

\def\grb{GRB\thinspace{980519}}
\def\vla{VLA\thinspace{J232221.5+771543}}
\def\ts{\thinspace}

\begin{document}
 
\title{\large \bf The Radio Afterglow From GRB\ts{980519}:
A Test of the Jet and Circumstellar Models}
 
\author{D. A. Frail\altaffilmark{1}, 
 S. R. Kulkarni\altaffilmark{2},
 R. Sari \altaffilmark{3},
 G. B. Taylor\altaffilmark{1}, 
 D. S. Shepherd\altaffilmark{2}, 
 J. S. Bloom\altaffilmark{2}, 
 C. H. Young\altaffilmark{4,1}, 
 L. Nicastro\altaffilmark{5}, 
 N. Masetti\altaffilmark{6}}

\altaffiltext{1}{National Radio Astronomy Observatory, P.~O.~Box O,
  Socorro, NM 87801}

\altaffiltext{2}{California Institute of Technology, 
Owens Valley Radio Observatory
105-24, Pasadena, CA 91125}

 \altaffiltext{3}{Theoretical Astrophysics, California Institute of
  Technology, MS 103-33, Pasadena, CA 91125.}

\altaffiltext{4}{Mississippi State University, Dept. of Physics and
  Astronomy, MS 39762}

\altaffiltext{5}{Istituto di Fisica Cosmica con Applicazioni
  all'Informatica, CNR, Via U. La Malfa 153, I-90146 Palermo, Italy}

\altaffiltext{6}{Istituto Tecnologie e Studio Radiazione Extraterrestre,
  CNR, Via Gobetti 101, I-40129 Bologna, Italy}

\begin{abstract}
 
  We present multi-frequency radio observations from the afterglow of
  \grb\ beginning 7.2 hours after the gamma-ray burst and ending 63
  days later. The fast decline in the optical and X-ray light curves
  for this burst has been interpreted either as afterglow emission
  originating from a collimated outflow -- a jet -- or the result of a
  blast wave propagating into a medium whose density is shaped by the
  wind of an evolved massive star.  These two models predict divergent
  behavior for the radio afterglow, and therefore, radio observations
  are capable, in principle, of discriminating between the two. We
  show that a wind model describes the subsequent evolution of the
  radio afterglow rather well. However, we see strong modulation of
  the light curve, which we interpret as diffractive scintillation.
  These variations prevent us from decisively rejecting the jet model.

\end{abstract}

\keywords{gamma rays:bursts -- radio continuum:general --
  cosmology:observations}

\clearpage
 
\section{Introduction}

Both geometry and environment can affect the evolution of gamma-ray
burst (GRB) afterglows (\cite{mrw98b}, \cite{pmr98}), giving us
insight into such important issues as the total energetics, the burst
event rate and the nature of GRB progenitors. The light curves and
spectra from some of the earliest afterglows showed good agreement
with fireball models of spherical blastwaves expanding into a
homogeneous medium (e.g. \cite{wrm97}, \cite{wax97b}).  However, over
the past year, we have come to recognize a class of GRBs whose
afterglows exhibited steeper than normal power-law decays (i.e.
$f_\nu\propto{t}^{-2}$).  Explanations for this behavior include
invoking a large value for the energy spectral index ($p \sim 4$) of
the radiation electrons (\cite{ggv+98b}), a jet-like geometry for the
relativistic shock (\cite{rho97b}, \cite{rho99}, \cite{sph99}), and a
class of inhomogeneous circumburst medium models, specifically those
shaped by the winds (mass loss) of progenitor stars (\cite{vie97},
\cite{cl99}, \cite{lc99}).

Empirical evidence connecting GRBs to the collapse of massive stars is
accumulating from studies of GRB 980425 (\cite{gvv+98},
\cite{kfw+98}), GRB 980326 (\cite{bkd+99}) and GRB 970228
(\cite{rei99}, \cite{gtv+99}).  Further evidence comes from the
studies of GRB environments, and includes the small measured GRB/host
galaxy offsets (\cite{bod+99}, \cite{bsp99}) and the growing number of
optically obscured afterglows such as GRB\ts{980329} (\cite{tfk+98})
and GRB\ts{970828} (Djorgovski {\em et al.} 1999)\nocite{dfk+99}.

The two models -- the jet and the wind-shaped circumburst medium (WCM)
-- are well motivated by sound physical and empirical considerations.
Jets have been invoked in part due to their universality, but also
specifically to account for the large inferred isotropic energy
release in some GRBs (e.g. GRB 990123; Kulkarni et al. 1999).  The
broad-band steepening of the optical and radio light curves seen in
GRB 990510 is explained very well by the jet model (\cite{sgk+99a},
\cite{hbf+99}). Likewise, if GRBs are the end points of massive stars
then it is inevitable that the circumburst medium will be
inhomogeneous and reflect the mass loss history of the progenitor
star.

\grb\ occupies a central position in the debate of jets versus WCM
models.  It is the second brightest GRB in the BeppoSAX sample, and
its X-ray and optical afterglow showed rapid fading, $f\propto t^{-2}$
(see \cite{hkpb99}).  Based on the optical and X-ray data Sari, Piran
\&\ Halpern (1999\nocite{sph99}) advocated a jet model whereas
Chevalier and Li (1999\nocite{cl99}) advocate a WCM model. We discuss
here the radio data in view of these two very different models.


\section{Observations}\label{sec:results}

\grb\ was detected on 1998 May 19.51 UT with the Wide Field Camera
(WFC) on board the BeppoSAX satellite (\cite{mhb+98}). A follow up
observation with the Narrow Field Instruments (NFI) some 10 hrs after
the burst detected a rapidly fading X-ray source (\cite{nac+98}) with
a power-law decay slope of $-$1.83$\pm$0.3 (\cite{naa+99}). Within the
WFC error circle, and coincident with the subsequent NFI location,
Jaunsen et al.  (1998\nocite{jha+98}) noted a fading optical
counterpart. Djorgovski et al.  (1998\nocite{dgk+98}) showed that the
optical transient (OT) exhibited a power-law decline with $\alpha\sim
-2$; here $f(t)\propto t^{\alpha}$. A summary of the X-ray and optical
afterglow light curves can be found in Nicastro et al.
(1999\nocite{naa+99}) and Halpern et al.  (1999\nocite{hkpb99}),
respectively.

We began observations of GRB 980519 with the Very Large Array (VLA)
some 7.2 hours after the initial $\gamma$-ray burst. The source was
clearly detected on May 22; hereafter, we will refer to the radio
afterglow as \vla.  A log of the VLA observations and a summary of the
results can be found in Table~\ref{tab:Table-VLA}.  Except on the
first day, when the bandwidth was halved in order to image the full
WFC error circle, a 100 MHz bandwidth was used and all four Stokes
parameters were measured.  As these observations were begun when the
VLA was in its most extended configuration and during the stormy
summer season, considerable care was taken to accurately track the
atmospheric phase variations across the array.  Each scan on the
source lasted three to five minutes and was bracketed on either side
with a nearby phase calibrator (J0017+815). As a check on the
integrity of the phase solutions a second phase calibrator (J2344+824)
was observed frequently. The flux scale was tied to one of the sources
J0137+331, J0542+498, or J1331+305.

In addition to the VLA observations, we observed the OT with the Owens
Valley Radio Observatory (OVRO) six-element array on the evening of
1998 May 21. \grb\ was observed for 6.5 hr at a central frequency of
100 GHz with a 2 GHz bandwidth.  The synthesized beam was
5\arcsec$\times${4}\arcsec; see Shepherd et al. (1998\nocite{sfkm98})
for the procedure used to analyze the data. The resulting map had an
rms noise level of 0.9 mJy beam$^{-1}$ and the flux at the position of
the OT was $-$1.2 mJy beam$^{-1}$. Thus we place an upper limit (mean
plus 3-$\sigma$) of 1.5 mJy beam$^{-1}$.

The light curve of \vla\ can be found in Figure~\ref{fig:rlight}. The
best fit position of \vla, determined by a Gaussian fit of the
combined 8.46 GHz data from May 22, May 24, June 2 and June 5 is
(epoch J2000) $\alpha$\ =\ $23^h22^m21.50^s$ ($\pm{0.02^s}$),
$\delta$\ =\ $+77^\circ15^\prime43.25^{\prime\prime}$
($\pm{0.06}^{\prime\prime}$).  The radio source coincides with the OT
within the errors of the optical astrometry (\cite{jha+98}).

\section{The Light Curve: Interstellar Scattering and 
  Scintillation}\label{sec:lcurve}

\begin{figure*} 
\centerline{\hbox{\psfig{figure=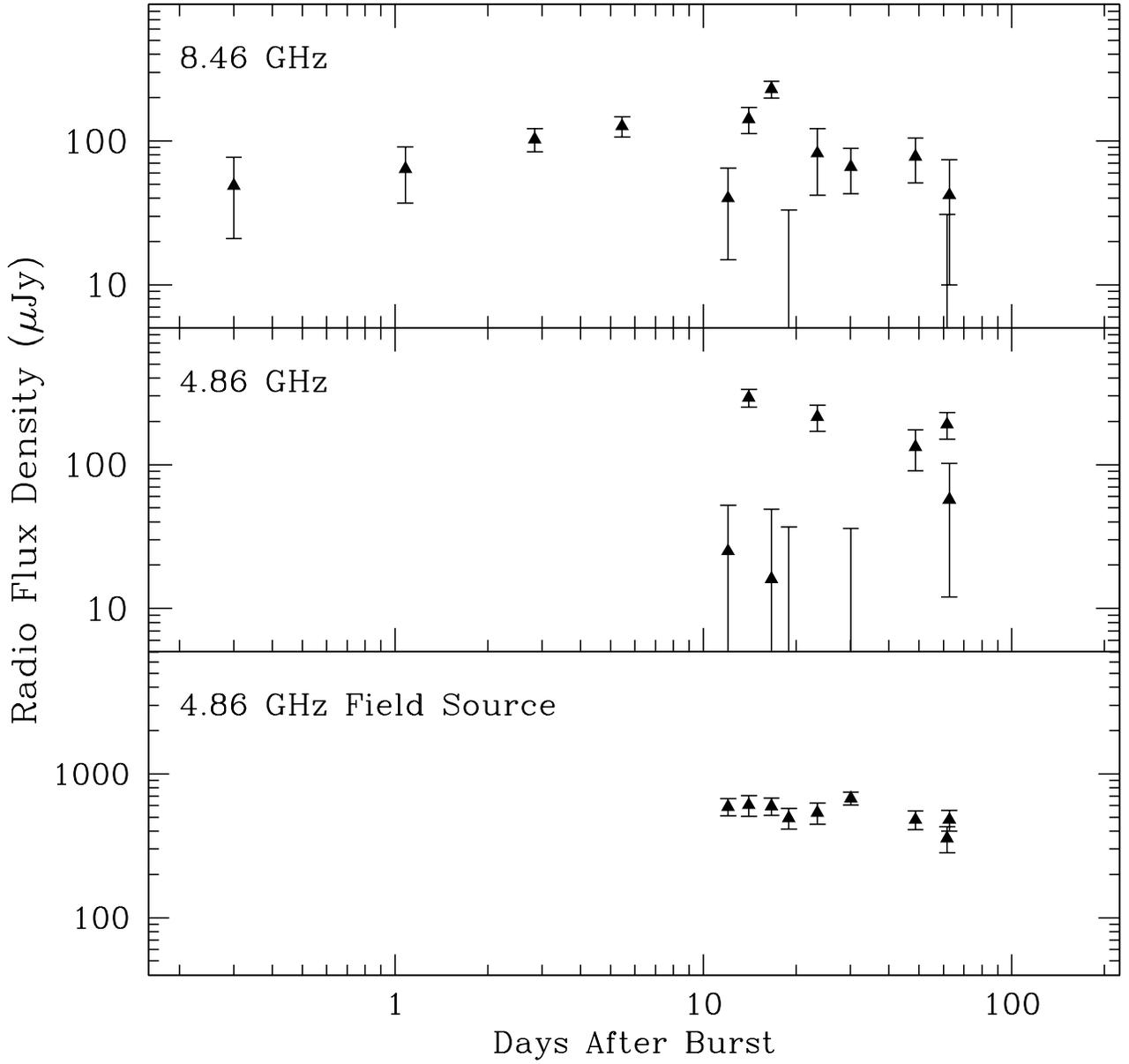,width=7.0in}}} 
\caption[]{Observed radio lightcurves at 8.46 GHz and 4.86 GHz. 
  The two top panels are light curves for the radio afterglow from
  \grb, while the bottom panel is the light curve for
  J232137.6+7715.0, a field source 2.5\arcmin\ from the GRB. Flux
  densities (and errors) are taken from Table~\ref{tab:Table-VLA}.
\label{fig:rlight}}
\end{figure*}

The light curve of \vla, with a gradual rise to a plateau followed by
a decline below detectability around day 60, is qualitatively similar
to that of many previously studied afterglows.  Even though the data
are sparse it is clear that there are some large amplitude variations
(e.g. around about June 7), especially at 4.86 GHz.  One could
attribute the variations to residual uncorrected phases caused by poor
weather.  Fortunately, there exists a field source J232137.6+7715.0,
some 2.5\arcmin\ from \grb. As can been seen from the two lower panels
in Fig. \ref{fig:rlight}, the flux of this source is stable and does
not show any suppression on those days when the afterglow was not
detected. Phase incoherence is a multiplicative error and the
stability of J232137.6+7715.0 is an empirical confirmation of the
robustness of our calibration procedure.  Parenthetically we note that
the field source is beyond the delay beam and is thus expected to
suffer more photometric errors than a source close to the phase
center.

We believe that the large variations seen in the light curve of \vla\ 
are real, and following Goodman (1997\nocite{goo97}) and Frail et al.
(1997\nocite{fkn+97}) we attribute these variations to interstellar
scattering and scintillation (ISS).  In particular, diffractive
scintillation can induce extreme intensity variations (see below) and
thereby account for the nulls in the light curve (Fig.
\ref{fig:rlight}).  For example, in the relatively short time window
of 12 to 19 days, we have four measurements with a mean of 103 $\mu$Jy
and standard deviation of 103 $\mu$Jy at 8.46 GHz, i.e. $100\%$
modulations. At 4.86 GHz we have mean of 80 $\mu$Jy and standard
deviation of 142 $\mu$Jy, i.e. $180\%$ modulations.  The statistical
fluctuations from thermal noise can explain about $30\%$ and $40\%$ of
the flux modulations at 8.46 GHz and 4.86 GHz, respectively. For
comparison, the field source had mean of 570 $\mu$Jy and standard
deviation of 52 $\mu$Jy, while the statistical fluctuations are 85
$\mu$Jy.  Therefore, the modulations in our data are likely dominated
by scintillation rather than by statistical noise. We now proceed with
the ISS hypothesis and use the observed variability to infer the
angular size of the source.

The strength of the scattering is reflected in the parameter SM, the
so-called scattering measure. In the direction towards the GRB
$(l,b)=(117.96^\circ,15.26^\circ$), using the formulation of Taylor
\&\ Cordes (1993\nocite{tc93}) we estimate ${\rm SM}=5.8\times
10^{-4}\,{\rm m}^{-20/3}\,{\rm kpc}$; for reference, we note that this
is a factor of two larger than towards GRB 970508 (\cite{fkn+97}).  We
have assumed that the typical pathlength to the scattering medium in
this direction is d$_{\rm scr}$=1.9 kpc. Using this SM and Goodman's
equations we expect to see strong scattering for frequencies below
$\nu_0=15$ GHz.  The corresponding Fresnel angle and Fresnel size at
$\nu_0$ are, $\theta_{F_0}=1.5\,\mu$arcsec and $R_{F_0}=4.2\times
10^{10}$ cm.

For $\nu>\nu_0$, the scattering is weak and the modulations scale as
$(\nu/\nu_0)^{-17/12}<1$. The fact that we observe order of unity
fluctuations at both 4.86 GHz and 8.46 GHz implies that $\nu_0\ge 8.5$
GHz, compatible with the estimate from the SM. For $\nu<\nu_0$ we will
see strong diffractive scintillation only if the size of the source,
$\theta_S<\theta_D=\theta_{F_0}(\nu/\nu_0)^{6/5}$.  The other ISS
parameters of interest are the decorrelation timescale (time for
significant changes in the detected flux), $t_{\rm
  RISS}=3.9{\rm\,hr}(\nu/\nu_0)^{6/5}(R_{F_0}/4.2\times
10^{10}{\rm\,cm}) /(v_{\perp}/30\,{\rm km\,s^{-1}})$ and the bandwidth
over which the diffractive ISS is decorrelated, $\Delta\nu = \nu_0
(\nu/\nu_0)^{22/5}$.  For $\nu_0\simeq{15}$ GHz the resulting
($\theta_D,t_{\rm diff}$, $\Delta\nu$) are (0.7$\,\mu$arcsec, 1.8 hr,
900 MHz) at 8.46 GHz, and (0.4$\,\mu$arcsec, 0.9 hr, 80 MHz) at 4.86
GHz. In calculating $t_{\rm RISS}$ we have assumed a transverse speed
of 30 km s$^{-1}$ for the scintillation pattern across the
line-of-sight.  These estimates are rough in the sense that the
estimate of SM (and hence $\nu_0$), for a given pathlength, can
typically fluctuate by a factor of a few.  Decreasing $\nu_0$ has the
effect of increasing $\theta_D$, $t_{\rm diff}$ and $\Delta\nu$.

These strong diffractive fluctuations can be suppressed under two
circumstances: (1) When the duration ($\Delta t$) and/or the bandwidth
($B$) of the observations exceed $t_{\rm RISS}$ and $\Delta\nu$,
respectively. The relevant figures of merit are given by
$n_1=\sqrt{B/\Delta\nu}$ and $n_2=\sqrt{\Delta{t}/t_{\rm RISS}}$.  (2)
If the source size, $\theta_S$, exceeds $\theta_D$. The figure of
merit here is $n_3=\theta_S/\theta_D$. If either $n_1$ or $n_2>1$ then
the observation will encompass more than one ``scintel'' (an island of
constructive interference), thereby suppressing the variations. A
large source will lower the modulation index to $n_3^{-1}$ and
increase the ISS timescale by the factor $n_3$ (\cite{nar93}).  The
fluctuations $\delta F$ of the observed flux $F$ in the regime of
diffractive scintillation will therefore be given by
\begin{equation}
\frac {\delta F} F = \min(1,n_1^{-1})\min(1,n_2^{-1})\min(1,n_3^{-1}).
\end{equation}

Our detection of strong diffractive scintillation at these frequencies
suggests that all three figures $n_1,n_2,n_3 \le 1$; otherwise the
fluctuations would be suppressed.  From Table~\ref{tab:Table-VLA} and
the estimate of the ISS parameters above we find $n_1(4.86\, {\rm
  GHz}) \sim 1$, $n_1(8.46\,{\rm GHz})\sim 0.1$ and $n_2(4.86\, {\rm
  GHz}) \sim 2$, $n_2(8.46\,{\rm GHz})\sim 1$. These estimates are
compatible with the constraint discussed above.  Interestingly, if
$\nu_0$were any larger than 15 GHz, the fluctuations would have been
suppressed since both $n_1$ and $n_2$ would have become larger than
unity at 4.86 GHz.  Our observations, therefore, limit $\nu_0$ to a
narrow range of values (${8-15}$ GHz) and likewise the estimates of
$\theta_D,t_{\rm diff}$, $\Delta\nu$.

Thus, given the uncertainties involved, the strong modulations seen at
both frequencies are quite reasonable.  Furthermore, the strong
modulations require that $n_3$ be close to unity i.e. the source must
be less than $\theta_D$.  Thus if our explanation that the strong
variation seen in the light curve is due to diffractive ISS is
correct, then the radio source must be less than $0.4\,\mu$arcsec -- a
very small size indeed.  Theoretical models predict a somewhat larger
size.  We will discuss this discrepancy in \S\ref{sec:discuss}.

The strong modulations caused by scintillation do not allow us to
accurately estimate the spectral slope in the radio band.  However,
averaging over all observations from day 12 on we find $\beta \cong
-0.45 \pm 0.6$ (where $f_\nu\propto\nu^\beta$). This large range is
compatible with both the early spectral index of $1/3$ expected at low
frequencies and with the late spectral index of $\sim -1$.  It is
inconsistent with the self-absorption spectral index of $2$.  The
source was never detected at 1.43 GHz (see Table~\ref{tab:Table-VLA}).
We conclude that \vla\ is an extremely compact ($<1\,\mu$arcsec)
source with a self absorption frequency $\nu_{ab}$ between 1.43 GHz
and 4.86 GHz.

\section{Discussion}\label{sec:discuss}

The intense interest in GRB 980519 primarily stems from the fact that
it was one of the first GRBs to be recognized to have a rapidly fading
afterglow (\cite{hkpb99}).  Two entirely different models were
proposed to account for the rapid fading: a jet expanding into a
homogeneous medium (\cite{sph99}) and a spherical explosion into a
wind-shaped circumburst medium (WCM; \cite{cl99}).  Both Sari et al.
and Chevalier \&\ Li used the same optical and X-ray data but ended up
favoring two radically different models.

The currently accepted view is that the afterglow emission arises in
the forward shock of relativistically moving material ejected from a
compact source (e.g. \cite{mr97a}, \cite{wax97a}).  It is assumed that
in this shock the electrons are accelerated to a power-law
distribution of energies (index $p$) above some minimum energy,
$\gamma_{m}m_ec^2$.  Furthermore, it is assumed that there is a
suitably strong magnetic field in the post-shock region.  The energy
density of the electrons and the magnetic field is assumed to scale
linearly with the energy density of the protons. Gyration of the
electrons in the magnetic field then results in a broad-band spectrum
characterized by three frequencies, $\nu_{ab}$, $\nu_m$ and $\nu_c$
and the flux at frequency $\nu_m$, $f_m$ (\cite{spn98}).  The slope
above $\nu_m$ is given by $f_\nu\propto\nu^{-(p-1)/2}$, and this
steepens by one half at $\nu_c$ due to radiative cooling. Below
$\nu_m$ the spectral slope is the classical $f_\nu\propto\nu^{1/3}$,
until it turns over below $\nu_{ab}$ due to synchrotron
self-absorption, for which $f_\nu\propto\nu^2$. The dynamics of the
expansion, the density distribution of the circumstellar matter and
the details of the energy injection govern the the temporal evolution
of these four parameters.  However, the basic shape of the above
spectrum does not change.  Thus the key to distinguishing between
competing models is through following the temporal evolution of the
spectrum.

Characterizing the broad-band spectrum by $f(\nu,t)\propto t^\alpha
\nu^\beta$, we note from Halpern et al.  (1999) in the first 1 -- 2
days after the burst the following: $\alpha_{\rm opt}=-2.05\pm$0.04,
$\beta_{\rm opt}=-1.20\pm 0.25$, $\alpha_{\rm X}=-1.83\pm 0.30$ and
$\beta_{\rm opt-X}=-1.05\pm 0.10$.  Within errors the X-ray afterglow
declines as rapidly as the better measured optical afterglow.
Likewise, within errors, the optical spectral index, $\beta_{\rm opt}$
is the same as the better measured optical-X-ray spectral index,
$\beta_{\rm opt-X}$.  Hereafter we will refer to $\alpha_{\rm opt}$
and $\beta_{\rm opt-X}$ by $\alpha$ and $\beta$.

Each model has a specific closure relation between $\alpha$ and
$\beta$.  We now consider each model in turn. (1) A spherical
explosion in a constant density medium would require $H=0$ or $H=1/2$
depending on whether the range of observing frequencies is below or
above $\nu_c$; here $S(\alpha,\beta)\equiv \alpha-3/2\beta$. The
measured $H=-0.48\pm 0.16$ rules out this model (Halpern et al. 1999,
Sari, Piran, \& Halpern 1999). (2) A jet model would need
$J(\alpha,\beta) \equiv \alpha -2\beta =-1,\, 0$ depending on whether
the range of observing frequencies is below or above $\nu_c$. We find
$J=0.05\pm 0.21$ which is seemingly more consistent with a jet
expanding into a constant density medium with both the optical and
X-ray frequencies below $\nu_c$. Sari et al. (1999) and Halpern et al.
(1999) favor a jet model but with $\nu_c$ between the optical and
X-ray bands, in order to explain $\beta$ with a standard value of $p
\sim 2.4$. The apparent inconsistency with the value of $J$ is
attributed, in this interpretation, to the theoretically very long
transition (\cite{pmr98}, \cite{rho99}, \cite{sph99}) to the jet
asymptotic power-law. The observed value of $\alpha$ may therefore be
less than the theoretical one, if measured on a finite time interval.
(3) An impulsive explosion in wind-shaped circumstellar model would
need $W(\alpha,\beta) =2\alpha-3\beta=-1,\, 1$ depending on whether
the range of observing frequencies is below or above $\nu_c$.  We find
$W=-0.95\pm 0.3$ which is consistent with this model provided that
both the optical and X-ray frequencies are below $\nu_c$. Chevalier
\&\ Li (1999) justify the WCM model on this basis. However, they do
note that their estimated $\nu_c$ on day 1 (when $\alpha$ and $\beta$
were measured) is between the optical and X-ray bands and argue that
given the scanty measurements and rough theory the inconsistency is
not particularly worrisome.

Thus both the jet model and the WCM model can account for the observed
optical and X-ray afterglow observations including the rapid decay.
In the WCM model, the rapid decline is due to the gas density in the
wind falling off as the inverse square of the distance between the
shock boundary and the burst location, together with a larger value of
the electron powerlaw index $p \sim 3$.  In the jet model, the
decrease is due to geometry. Due to relativistic ``beaming'', only a
small portion of the shocked gas is visible to the observer, a solid
angle $\Gamma^{-2}$ where $\Gamma$ is the bulk Lorentz factor of the
shocked gas.  When $\Gamma$ falls below $\theta_J^{-1}$, the inverse
of the opening angle of the jet, the observer will notice the finite
edge of the emitting surface, thereby leading to a deficit of emission
and thus a steeper than normal decline. Moreover, at about the same
time, the jet will begin to expand sidewise, further enhancing the
decline.

We now discuss how radio observations can help resolve the ambiguity
of the choice of models. For the sake of completeness, we also include
a discussion of the simplest afterglow model (spherical explosion
in a homogeneous medium).

\begin{figure*}
  \centerline{\hbox{\psfig{figure=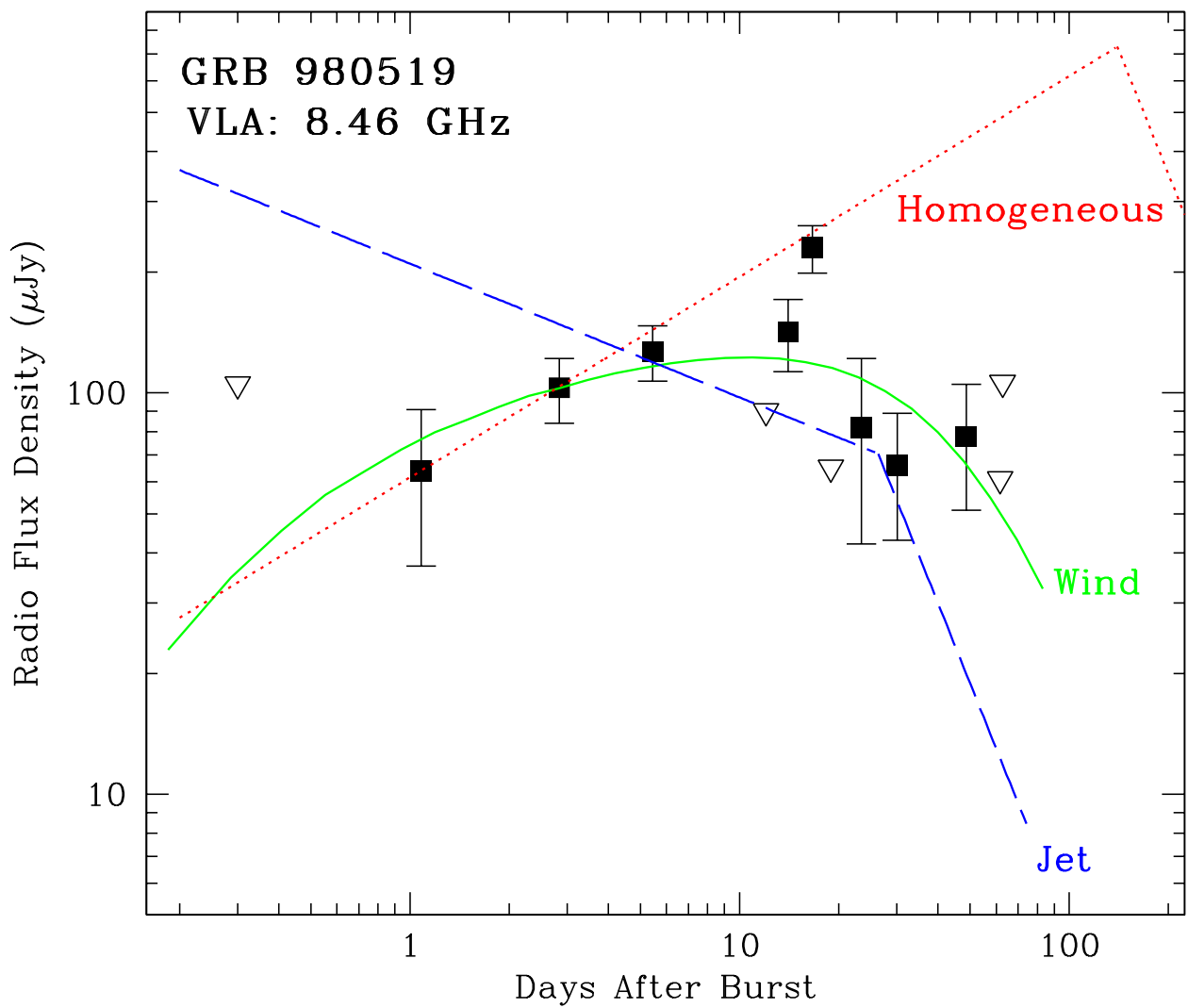,width=7.0in}}}
\caption[]{Observed and model lightcurves at 8.46 GHz. Detections 
  are indicated by the filled squares. Upper limits for the
  non-detections (open triangles), are plotted as the measured flux at
  the position of the radio transient plus twice the rms noise.  Three
  different model predictions are plotted. The circumstellar model
  ``Wind'' is taken directly from Chevalier \& Li (1999) with no
  modifications. The basic adiabatic, forward shock model with
  spherical expansion into a homogeneous medium is indicated as
  ``Homogeneous''. The predicted behavior for the jet model of Sari et
  al.~(1999) is given by ``Jet''.  See text for additional details.
  \label{fig:model}}

\end{figure*}

\noindent{\bf Spherical, homogeneous model.} In this model, 
$\nu_{ab}\propto{t}^0$, $\nu_m\propto{t}^{-3/2}$ and $f_m=t^0$.  As
discussed in \S\ref{sec:lcurve}, $\nu_{ab}<$8.46 GHz$<\nu_m$; this is
certainly a secure statement two days after the burst. In this
frequency range, we expect $f(t)\propto t^{1/2}$ until $\nu_m$ reaches
8.46 GHz; we estimate (in the framework of this model) this will
happen around 140 d. This estimate relies on a knowledge of the basic
spectral shape of the afterglow (\cite{spn98}) and the near
simultaneous optical and radio data on 1998 May 28.59 UT to derive
$\nu_m$ and $f_m$ at this time and evolve it forward until
$\nu_m$=8.46 GHz. Thereafter the radio flux will decay in a manner
similar to the optical and X-ray flux. As can be seen from
Figure~\ref{fig:model} the predictions are clearly inconsistent with
the observations. Even the strong modulations produced by the ISS are
unable to resolve this discrepancy (see below). Thus we can reject the
spherical model in an entirely independent way from the closure
method.

\noindent{\bf Jet Model.} We begin by noting that the steep optical 
decline had already started at the time of the first afterglow
observation ($t_1$) -- the I-band observation by Jaunsen et
al.~(1998\nocite{jha+98}) about 7.5 hr after the burst. Thus the epoch
at which the jet-like geometry of the emitting surface becomes
apparent to the observer, $t_J$, is less than $t_1$.  The epoch of the
I-band observations is close to our first radio observation (see
Table~\ref{tab:Table-VLA}). In this ``jet-dominated'' regime the radio
afterglow emission is $f(\nu_{ab}<\nu<\nu_m) \propto t^{-1/3}$ and
$f(\nu<\nu_{ab}) \propto t^0$ (Sari et al. 1999).  However, in
\S\ref{sec:results} we argued that \vla\ is an absorbed source with
$\nu_{ab}$ between 1.43 and 4.86 GHz. The self-absorption frequency
evolves slowly with time, $\nu_{ab}\propto t^{-1/5}$. Thus we expect
8.46 GHz to be in the optically thin regime and therefore the flux at
this frequency should decrease as $\propto t^{-1/3}$.  A $t^{-1/3}$
fit to the data, does not give a reasonable $\chi^2$.  However, the
errors in the flux density (as quoted in Table~\ref{tab:Table-VLA})
may be underestimating the real uncertainties for a strongly
scintillating source. If we fit the data, allowing for strong
diffractive scintillation in the measurement uncertainties (i.e.
$\sigma=\bar{S}$) then a reasonable solution can be found
S=210\,$t_d^{-1/3}$ $\mu$Jy ($\chi^2$=6 with 13 dof), where $t_d$ is
time in days.  The sharp break in the light curve at $t_m=26$ days
corresponds to $\nu_m$=8.46 GHz and was derived by evolving the
spectrum forward from $t=1.08$ days (with $\beta$=1/3 for $\nu<\nu_m$
and $\beta=\beta_{\rm opt-X}$ for $\nu>\nu_m$), when high quality
optical and radio data existed.


\noindent{\bf Wind-shaped Circumstellar Medium Model.}
Using the contemporaneous X-ray, optical and radio data taken up to
three days after the burst, Chevalier \& Li (1999) fit the
circumstellar model to derive the fireball parameters (total energy,
energy power law index, fraction of energy in electrons and magnetic
fields) and the wind properties, for an assumed redshift of $z$=1. In
their derivation, the 8.46 GHz emission initially originates below
$\nu_{ab}$ and $\nu_{m}$ but the subsequent evolution moves them into
this band as $\nu_{ab}\propto{t}^{-3/5}$, $\nu_m\propto{t}^{-3/2}$ and
$f_m\propto{t}^{-1/2}$. Thus in the circumstellar model one predicts
an initial rise of the radio flux as the source opacity decreases,
followed by a plateau when $\nu_{ab}<$8.46 GHz$<\nu_m$ and then a
steep decay when $\nu_m<$8.46 GHz. We have plotted the model fit of
Chevalier \& Li (1999) (labeled ``Wind'') in Fig.~\ref{fig:model}.  We
stress that the model was derived on the first three days of afterglow
data, spanning 8 orders of magnitude in frequency, and yet without any
further adjustments their model fits the next 60 days of VLA
measurements remarkably well.

As we have noted in the previous section, the angular size of the
afterglow is less than $1\,\mu$arcsec even after about 15 days.  In
the three models, spherical, wind and jet, the afterglow angular size
at that time is given by
\begin{equation}
\theta_{Sphere}=
2.8\,\mu{\rm arcsec} \left( \frac {1+z} 2 \right)^{-5/8} D_{A,28}^{-1}
E_{52}^{1/8} n_i^{-1/8} (t/15\,{\rm days})^{5/8}
\end{equation}
\begin{equation}
\theta_{Wind}=
2.2\,\mu{\rm arcsec} \left( \frac {1+z} 2 \right)^{-3/4} D_{A,28}^{-1}
E_{52}^{1/4} A_\star^{-1/4} (t/15\,{\rm days})^{3/4} 
\end{equation}
\begin{equation}
\theta_{J}=
1.7\,\mu{\rm arcsec} \left( \frac {1+z} 2 \right)^{-5/8} D_{A,28}^{-1}
E_{52}^{1/8} n_i^{-1/8} (t_{J}/8\,{\rm hr})^{1/8} (t/15\,{\rm days})^{1/2}.
\end{equation}
Here, $D_{A,28}$ is the distance in units of 10$^{28}$ cm, $E_{52}$ is
the inferred ``isotropic'' energy released in the explosion in units
of $10^{52}$ergs, $n_i$ is the surrounding density in the constant
density case. $A_\star$ characterizes the wind density as
$\rho(R)=5\times 10^{11} A_\star R_{\rm cm}^{-2}$ g cm$^{-3}$, with as
a $R_{\rm cm}$ the wind radius in cm.  These sizes are quite similar
to each other. This stems from the very low dependence of the size as
a function of density and from the fact that the density in front of
the shock, at a time of 15 days, in a wind model with typical
parameters is not very different from regular ISM densities.  It can
be seen that the small inferred size from the scintillation is
challenging for all these models. It requires either a dense wind, or
a jet with very small opening angle, that begins to spread at a very
early time.  It is hard to reduce the size below $1\,\mu$arcsec for
any reasonable choice of the parameters in the spherical constant
density model. The agreement with scintillation theory would improve
if either SM or the effective distance of the scattering screen
$d_{\rm scr}$ were lower by a factor of two. Along specific lines of
sight the uncertainty in either of these quantities could vary by this
amount due to the non-uniform nature of the ionized medium.  In such a
case the inferred size of $0.4\,\mu$arcsec would increase by a factor
of 2.5, enabling both jet and wind models to fit the data.

\section{Conclusions}

The observed steep decline in the optical and X-ray light curves of
the afterglow from \grb\ lends itself to two equally compelling
hypotheses, the first being that the afterglow emission originates
from a collimated outflow (or jet), and the second, that the emission
is the result of a blast wave propagating into a medium whose density
is shaped by the wind of an evolved massive star.  These competing
models predict divergent behavior for the evolution of the radio
emission. In the jet model the radio flux is expected to decrease
continuously after the jet edge becomes visible. In contrast, the
circumstellar model, specifically the Wolf-Rayet wind model of
Chevalier \& Li (1999), predicts a linear rise, followed by a broad
plateau and a late decay in the radio light curve.

In this case, scintillation, low-signal-to-noise and sparse data at
early times, has limited us from decisively choosing either model.
The influence of ISS could be reduced, in future afterglow
observations, if one has more measurements at early times (the first
two days is where the models differ the most). Such observations,
separated by a few hours (a time which is larger than $t_{ISS}$),
could be averaged to produce a more solid estimate of the emitted
flux, free of large modulations. Nevertheless, this result illustrates
the unique diagnostics provided by radio observations and its
potential for unraveling the origin of GRBs. Previous examples include
the demonstration of jets in GRB 990510 (\cite{hbf+99}) and the
discovery of a reverse shock in GRB 990123 (\cite{sp99b},
\cite{kfs+99}).

\acknowledgements

DAF thanks Roger Chevalier and Zhi-Yun Li for providing their model
light curves. SRK thanks Jim Cordes for discussions related to ISS.
The Very Large Array (VLA) is operated by the National Radio Astronomy
Observatory, a facility of the National Science Foundation operated
under cooperative agreement by Associated Universities, Inc.  SRK's
research is supported by grants from NSF and NASA.  RS is supported by
the Sherman Fairchild Foundation.

\clearpage
  
 

\clearpage
 
\newpage 
\begin{deluxetable}{lrcrrrr}
\tabcolsep0in\footnotesize
\tablewidth{\hsize}
\tablecaption{VLA Observations of \grb\ \label{tab:Table-VLA}}
\tablehead {
\colhead {Epoch}      &
\colhead {$\Delta t$} &
\colhead {TOS} &
\colhead {S$_{8.46}\pm\sigma$}  &
\colhead {S$_{4.86}\pm\sigma$}  &
\colhead {S$_{field}\pm\sigma$} &
\colhead {S$_{1.43}\pm\sigma$}  \\
\colhead {(UT)}      &
\colhead {(days)} &
\colhead {(min)} &
\colhead {($\mu$Jy)} &
\colhead {($\mu$Jy)} &
\colhead {($\mu$Jy)} &
\colhead {($\mu$Jy)}
}
\startdata

1998 May 19.81  & 0.30  & 99  & 49$\pm$28  & \omit & \omit & \omit \nl
1998 May 20.59  & 1.08  & 54  & 64$\pm$27  & \omit & \omit & \omit \nl
1998 May 22.35  & 2.84  & 112 & 103$\pm$19 & \omit & \omit & \omit \nl
1998 May 24.96  & 5.45  & 112 & 127$\pm$20 & \omit & \omit & \omit \nl
1998 May 31.54  & 12.03 & 132 & 40$\pm$25  &  25$\pm$27   & 591$\pm$80  & \omit \nl
1998 Jun. 02.56 & 14.05 & 125 & 142$\pm$29 & 292$\pm$41   & 607$\pm$100 & \omit \nl
1998 Jun. 05.14 & 16.63 & 106 & 230$\pm$31 & 16$\pm$33    & 596$\pm$80 & \omit \nl
1998 Jun. 07.43 & 18.92 & 183 & 1.2$\pm$32 & $-$13$\pm$37 & 493$\pm$82 & $-$4.5$\pm$34 \nl
1998 Jun. 11.94 & 23.43 & 69  &  82$\pm$40 & 215$\pm$44   & 537$\pm$90 & \omit \nl
1998 Jun. 18.58 & 30.07 & 78  &  66$\pm$23 &  5$\pm$31    & 676$\pm$70 & \omit\nl
1998 Jul. 07.26 & 48.75 & 100 &  78$\pm$27 &  133$\pm$42  & 480$\pm$71 & \omit \nl
1998 Jul. 20.23 & 61.72 & 124 & 0.9$\pm$30 & 190$\pm$40   & 356$\pm$73 & 15$\pm$42 \nl
1998 Jul. 21.25 & 62.74 & 94  &  42$\pm$32 &  57$\pm$45   & 480$\pm$79 & $-$35$\pm$58 \nl
\enddata
\tablecomments{The columns are (left to right), (1) UT date of the
  start of each observation, (2) time elapsed since the GRB 980519
  event, (3) total duration of the observing run, (4) 8.46 GHz flux
  density of the radio transient (RT), with the error given as the
  root mean square flux density, (5) 4.86 GHz flux density of the RT,
  (6) The integrated 4.86 GHz flux density of J232137.6+7715.0, a
  field source 2.5\arcmin\ from the GRB, and (7) 1.43 GHz flux density
  of the RT.}
\end{deluxetable}

\clearpage 

\clearpage 
\end{document}